# Combined Method Using Bohmian Mechanics and Quantum Phase Space Representation

Dmytro Babyuk

*Department of Chemistry, University of Nevada, Reno, Reno Nevada 89557*

A combined method for analyzing quantum dynamical equations which uses the Bohmian mechanics and the quantum phase space representation is proposed. It is based on a presentation of the wave function in phase space in a polar form. The combined method is very effective for comparison of classical and quantum dynamics for polynomial potentials of order no higher than second. Application for the harmonic oscillator and quartic double well potential is performed. Quantum fluxes and particle energy are derived by means of the proposed method.

## Introduction

Governing processes in the microworld, quantum mechanics appears as an exact theory which provides full information about microsystem. But even at present time of powerful calculating-machines it cannot be employed successfully for the realistic many-dimensional systems due to numerical difficulties arisen from dealing with matrices of a big size. Therefore classical-quantum correspondence is one of the central problem in modern physics. In current paper we consider only two approaches widely used for its handling.

The first approach is a Bohmian mechanics (BM) or hydrodynamic formulation of quantum mechanics. According to the BM the wave function is represented in a polar form [1]. Then the Schroedinger equation decomposes into two parts: an analogue of the classical Hamilton-Jacobi (HJ) equation and probability conservation equation. Quantum effects are included into an additional term in the HJ equation, called quantum potential. The presence of the latter gives rise to a problem associated with numerical difficulties of its calculation.

The second approach is a quantum phase space representation (QPSR). Due to the uncertainty principle there exists a variety of them. It is hard to pick out one as ideal because each approach has its advantages and disadvantages. In current work we do not pursue an aim to characterize them and will consider only one, proposed by Go.Torres-Vega and Frederick [2]. The main idea of it is to formulate a definition of coordinate $\hat{Q}$ and momentum $\hat{P}$ operators in phase space in a way the uncertainty principle is satisfied. From that operators the Hamiltonian operator is constructed. Thus the time-dependent Schroedinger equation in phase space which governs the wave function dependent on momentum and coordinate simultaneously can be derived. Both BM and QPSR are as an analytical tool for learning of classical-quantum correspondence rather than a simulation tool. However, the quantum trajectory method based on direct integration of the BM equations recently developed by Wyatt group has been applied to several problems [3-5].

## CM formulation

Despite of diversity of both methods we propose a combined method (CM) based on the BM and QPSR. The Schroedinger equation in phase space representation [2]

$$i\frac{\partial \Psi(p,q,t)}{\partial t} = \frac{1}{2}\left(\frac{p}{2} - i\frac{\partial}{\partial q}\right)^2 \Psi(p,q,t) + \hat{U}\left(\frac{q}{2} + i\frac{\partial}{\partial p}\right)\Psi(p,q,t) \quad (1)$$

can be written with wave function in polar form

$$\Psi(p,q,t) \to \Phi(p,q,t) = R(p,q,t)e^{iS(p,q,t)}, \quad (2)$$

where $R$ and $S$ are real function and they are related with $\Psi$ by

$$R(p,q,t) = \sqrt{\text{Re}(\Psi(p,q,t))^2 + \text{Im}(\Psi(p,q,t))^2}$$

$$S(p,q,t) = \arg(\Psi(p,q,t))$$

The standard separation of real and imaginary variables leads to two equations

$$-\frac{\partial S}{\partial t} = \frac{1}{2}P^2 - \frac{1}{2R}\frac{\partial^2 R}{\partial q^2} + V(P,Q,R,t) \quad (3)$$

$$\frac{\partial R^2}{\partial t} = -\frac{\partial}{\partial q}\left[R^2 P\right] + \frac{\partial}{\partial p}G(P,Q,R,t), \quad (4)$$

where

$$P(p,q,t) = \left[\frac{p}{2} + \frac{\partial S}{\partial q}\right], \quad Q(p,q,t) = \left[\frac{q}{2} - \frac{\partial S}{\partial p}\right] \quad (5)$$

The form of $V(P,Q,R,t)$ and $G(P,Q,R,t)$ is determined by specifying the potential. Equation (3) can be considered as an analogue of the classical HJ equation and (4) as a quantum Liouville equation. Unlike BM, they include additional terms due to dependence of amplitude $R$ and phase $S$ on $p$ and $q$.

It can be shown that for any polynomial potential parameters $V$ and $G$ are reduced to a form containing only $P$, $Q$ (not $p,q$) and their derivatives with respect to $p$ and $q$. Moreover, as concluded from (4)

$$Jq(p,q,t) = R^2 P, \qquad Jp(p,q,t) = -G$$

i.e., the quantum fluxes are derived automatically without referring to the quantum Liouville equation as in the QPSR [2].

We have to admit that our approach suffers from an obvious shortcoming associated with presence of high order momentum derivatives in equations. The similar inconvenience is typical for momentum representation in quantum mechanics.

### Harmonic oscillator potential

The CM is especially effective in description for polynomial potentials of order no higher than second. We demonstrate it for the harmonic potential with $\omega = 1$. In this case (3)-(4) become

$$-\frac{\partial S}{\partial t} = \frac{1}{2}P^2 - \frac{1}{2R}\frac{\partial^2 R}{\partial q^2} + \frac{1}{2}Q^2 - \frac{1}{2R}\frac{\partial^2 R}{\partial p^2} \qquad (6)$$

$$\frac{\partial R^2}{\partial t} = -\frac{\partial}{\partial q}[R^2 P] + \frac{\partial}{\partial p}[R^2 Q] \qquad (7)$$

Quantum fluxes are

$$Jq(p,q,t) = R^2 P, \qquad Jp(p,q,t) = -R^2 Q \qquad (8)$$

They remind the classical ones with difference that instead $p$ and $q$ are $P$ and $Q$, respectively. According to (5) and (8) the quantum trajectories must coincide with classical ones if $\frac{\partial S}{\partial q} = \frac{p}{2}$ and $\frac{\partial S}{\partial p} = -\frac{q}{2}$. It will be shown below that it is true for trajectories starting from the center of the Gaussian wave packet.

It is worth emphasizing that $P$ and $Q$ coincide with local expectation values of $p$ and $q$ [1]

$$P_{local} = \text{Re}\left[\frac{\hat{P}\Psi}{\Psi}\right] = \text{Re}\left[\frac{p}{2} - i\frac{1}{\Psi}\frac{\partial \Psi}{\partial q}\right] = \left[\frac{p}{2} + \frac{\partial S}{\partial q}\right] = P(p,q,t)$$

$$Q_{local} = \text{Re}\left[\frac{\hat{Q}\Psi}{\Psi}\right] = \text{Re}\left[\frac{q}{2} + i\frac{1}{\Psi}\frac{\partial \Psi}{\partial p}\right] = \left[\frac{q}{2} - \frac{\partial S}{\partial p}\right] = Q(p,q,t)$$

Defining a new variable

$$W(p,q,t) = -\frac{1}{2R}\frac{\partial^2 R}{\partial q^2} - \frac{1}{2R}\frac{\partial^2 R}{\partial p^2} \qquad (9)$$

(6) is similar to the BM equation

$$-\frac{\partial S}{\partial t} = \frac{1}{2}P^2 + \frac{1}{2}Q^2 + W \qquad (10)$$

One might conclude from (10) that W is a quantum potential. We show that it is untrue. Differentiating (10) with respect to $p$ and using (8) gives

$$\frac{dQ}{dt} = P + \frac{1}{2}\frac{\partial W}{\partial P} \qquad (11)$$

Performing similar procedures with (10) but differentiating with respect to $q$ gives

$$\frac{dP}{dt} = -Q - \frac{1}{2}\frac{\partial W}{\partial Q} \qquad (12)$$

Without the second term on the right side of (11) and (12) would be the equations of classical motion in terms of $P$, $Q$ for the harmonic oscillator potential. It is clear now $W$ is not a quantum potential as in the BM because its derivatives are included in both equations.

An interesting point is to study the behavior of quantum trajectories. The Gaussian wave packet is the best choice as initial state because there exists an exact analytical solution. According to [2]

$$\Psi(p,q,t) = \frac{1}{\sqrt{2\pi}}\exp\left[-\frac{1}{4}(q-q_c(t))^2 - \frac{1}{4}(p-p_c(t))^2 + \frac{i}{2}(qp_c(t) - pq_c(t) - t)\right] \qquad (13)$$

where

$$q_c(t) = q_c(0)\cos(t) + p_c(0)\sin(t)$$

$$p_c(t) = p_c(0)\cos(t) - q_c(0)\sin(t)$$

Comparison of (13) with (2) gives the following amplitude and phase

$$R(p,q,t) = \frac{1}{\sqrt{2\pi}}\exp\left[-\frac{1}{4}(q-q_c(t))^2 - \frac{1}{4}(p-p_c(t))^2\right] \qquad (14)$$

$$S(p,q,t) = \frac{1}{2}[qp_c(t) - pq_c(t) - t]$$

It is easy to see that

$$P = \frac{p}{2} + \frac{\partial S}{\partial q} = \frac{1}{2}(p + p_c(t)) \quad Q = \frac{q}{2} - \frac{\partial S}{\partial p} = \frac{1}{2}(q + q_c(t)) \quad (15)$$

$$W(p,q,t) = -\frac{1}{4}\left[\frac{1}{2}(q - q_c(t))^2 + \frac{1}{2}(p - p_c(t))^2 - 2\right] \quad (16)$$

Substituting (16) into (11)-(12) and taking into account (15) gives

$$\frac{dq}{dt} = P = \frac{1}{2}(p + p_c(t)) \quad (17)$$

$$\frac{dp}{dt} = -Q = -\frac{1}{2}(q + q_c(t)) \quad (18)$$

We derived the equations for quantum trajectories as in [2]. The solution of this system is known

$$p(t) = [p(0) - p_c(0)]\cos\left(\frac{t}{2}\right) - [q(0) - q_c(0)]\sin\left(\frac{t}{2}\right) + p_c(t)$$

$$q(t) = [q(0) - q_c(0)]\cos\left(\frac{t}{2}\right) + [p(0) - p_c(0)]\sin\left(\frac{t}{2}\right) + q_c(t) \quad (19)$$

As seen, from (15), (17)-(18) classical and quantum trajectories coincide when starting point is the Gaussian center. $\frac{\partial S}{\partial q}$ and $\frac{\partial S}{\partial p}$ are more complicated for higher order potentials and the coincidence between trajectories cannot take place.

In contrast to the QPSR, the CM allows to introduce an energy of quantum particle which is exploring phase space. By analogy with the HJ equation the energy is given by

$$E(t) = -\frac{\partial S}{\partial t} \quad (20)$$

$$= \frac{1}{2}[q_c(0)^2 + p_c(0)^2 + 1 + (q_c(0)q(0) + p_c(0)p(0) - q_c(0)^2 - p_c(0)^2)\cos\left(\frac{t}{2}\right) + (p_c(0)q(0) - q_c(0)p(0))\sin\left(\frac{t}{2}\right)]$$

The figure shows an influence of initial condition on energy evolution.

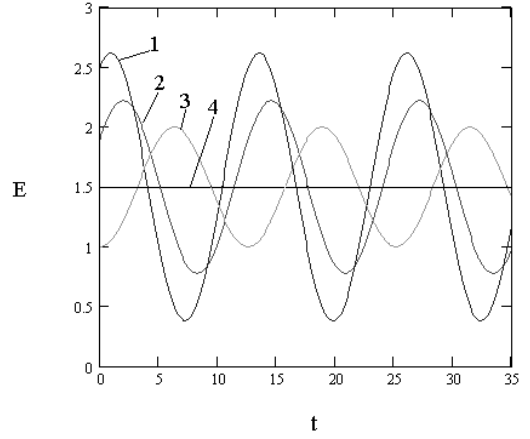

**FIG**. Energy dependence on time for different trajectories: 1) $p(0)=1.5$, $q(0)=2.5$; 2) $p(0)=0.8$, $q(0)=2$; 3) $p(0)=0.5$, $q(0)=0.5$; 4) $p(0)=1$, $q(0)=1$. Initial condition for all cases is $p_c(0)=1$, $q_c(0)=1$.

If the initial condition of quantum particle coincides with the Gaussian center, then its energy does not depend on time and the motion is like classical one

$$E(t) = -\frac{\partial S}{\partial t} = \frac{1}{2}[q_c(0)^2 + p_c(0)^2 + 1] = E_{cl}$$

Otherwise, the energy is a periodic function of time with period $4\pi$ but its average value remains equal to classical one

### Double well potential

As pointed out above, there is no such obvious analogy between classical and quantum equations for higher order potential. As an example, we consider the quartic potential with double well. For this case the CM can be useful in calculation of quantum fluxes and particle energy. If the potential is determined by

$$U(\hat{Q}) = \alpha\hat{Q}^4 - \beta\hat{Q}^2 + E_0 = \alpha\left(\frac{q}{2} + i\frac{\partial}{\partial p}\right)^4 - \beta\left(\frac{q}{2} + i\frac{\partial}{\partial p}\right)^2 + E_0$$

with constant $\alpha$, $\beta$ and $E_0$, then dynamical equations for it are written in the form

$$-\frac{\partial S}{\partial t} = \frac{1}{2}P^2 - \frac{1}{2R}\frac{\partial^2 R}{\partial q^2} + \alpha A(p,q,t) - \beta\left(Q^2 - \frac{\partial^2 R}{\partial p^2}\right) + E_0 \quad (21)$$

$$\frac{\partial R^2}{\partial t} = -\frac{\partial}{\partial q}Jq(p,q,t) + \frac{\partial}{\partial p}Jp(p,q,t) \quad (22)$$

where

$$A(p,q,t) = Q^4 + \frac{1}{R}\frac{\partial^4 R}{\partial p^4} - \frac{6Q^2}{R}\frac{\partial^2 R}{\partial p^2} - 4Q\left[\frac{\partial^2 Q}{\partial p^2} + \frac{3}{R}\frac{\partial R}{\partial p}\frac{\partial Q}{\partial p}\right] - 3\left(\frac{\partial Q}{\partial p}\right)^2$$

$$Jq(p,q,t) = R^2 P$$

$$Jp(p,q,t) = -2\left[\alpha\left(2R^2 Q^3 - B(p,q,t)\right) - \beta R^2 Q\right]$$

$$B(p,q,t) = \frac{\partial}{\partial p}\left(R^2 \frac{\partial Q}{\partial p}\right) + 6RQ\frac{\partial^2 R}{\partial p^2} - Q\frac{\partial^2}{\partial p^2}(R^2)$$

(23)

It is not possible to derive from (21)-(22) equations similar to (11)-(12) as in case of the harmonic oscillator potential. But from (23) one can still construct equations for quantum trajectories and integrate them numerically. The tunneling path between the wells can be studied in this way [6].

**Summary**

In this paper, we have introduced a method which combines the Bohmian mechanics and the quantum phase space representation. Being combined, it absorbs some advantages and disadvantages of both of them as well. The CM is a better tool for the classical-quantum correspondence than the BM and QPSR individually. It generates the quantum Liouville equation automatically and allows to introduce the quantum particle energy as negative time derivative of phase. An obvious drawback of the CM is an appearance of cumbersome equations for high order potential. However, using the similar routine from the BM which consist in numerical integration of the Schroedinger equation and then recovering the amplitude and phase, one can obtain the quantum fluxes required to derive quantum trajectories in phase space.

**Acknowledgement**

The author thanks John H.Frederick for helpful discussions.